\documentclass[]{mn2e}
\usepackage{amssymb}
\usepackage{graphicx}
\usepackage{amsmath}
\newif\ifAMStwofonts

\def\ARAA{ARA\&A}

\def\AnA{A\&A}

\def\ee #1 {\times 10^{#1}}
\def\ut #1 #2 { \, \rmn{#1}^{#2}}
\def\u #1 { \, \rmn{#1}}

\def\half{{\textstyle \frac{1}{2}}}
\def\thalf{{\textstyle{ 3\over 2}}}
\let\grad=\nabla
\def\cross{\bmath{\times}}
\def\curl #1 {\grad \cross #1}
\def\div #1 {\grad \cdot #1}

\def\etaA{\eta_{\rm A}}
\def\etaO{\eta_{\rm O}}
\def\etaH{\eta_{\rm H}}

\def\e{\bmath{e}}
\def\v{\bmath{v}}

\def\B{\bmath{B}}
\def\E{\bmath{E}}            
\def\Bh{\bmath{\hat{B}}}
\def\fh{\bmath{\hat{\phi}}}  
\def\rh{\bmath{\hat{r}}}     

\def\vr{v_{r}}
\def\vf{v_{\phi}}            
\def\vk{v_{K}}               

\def\Epa{\bmath{E'_\parallel}}  
\def\Epe{\bmath{E'_\perp}}  
\def\J{\bmath{J}}

\def\dE{\bmath{\delta\E}}
\def\dB{\bmath{\delta\B}}

\newcommand{\delt} [1] {\frac{\partial #1}{\partial t}}

\input epsf

\title{Magnetorotational instability in protoplanetary discs: The effect of dust grains}
\author[R. Salmeron and M. Wardle]
       {Raquel Salmeron$ ^{1,2} $
\& Mark Wardle$ ^3 $ \\
$ ^1 $Planetary Science Institute, Research School of Astronomy \& Astrophysics and 
Research School of 
Earth Sciences, \\ Australian National University,
Canberra ACT 2611, Australia \\
$ ^2 $Department of Astronomy \& Astrophysics, The University of Chicago, Chicago IL 60637, USA\\
$ ^3 $Physics Department, Macquarie University, Sydney NSW 2109, Australia}

\date{2007 March 20}
\pagerange{\pageref{firstpage}--\pageref{lastpage}}
\pubyear{2007}
\begin{document}
\maketitle
\label{firstpage}
\begin{abstract}
We investigate the linear growth and vertical structure of the 
magnetorotational instability (MRI) in weakly ionised, stratified protoplanetary 
discs. The magnetic field is initially 
vertical and dust grains are assumed to be well mixed with the gas over the entire vertical dimension of the disc.
For simplicity, all the grains are assumed to have the same radius ($a = 0.1$, $1$ 
or $3\  \mu$m) and constitute a constant fraction (1 \%) of the total mass of the gas. Solutions are obtained at representative radial locations ($R = 5$ and $10
$ AU) from the central protostar for a minimum-mass solar nebula model and different choices of the initial magnetic field strength, 
configuration of the diffusivity tensor and grain sizes. 

We find that when no grain are present, or they are $\gtrsim 1\ \mu$m in radius, the midplane of the disc remains magnetically coupled for field strengths up to a few gauss at both radii. In contrast, when a population of small grains ($a = 0.1 \mu$m) is mixed with the gas, the section of the disc within two tidal scaleheights from the midplane is magnetically inactive and only magnetic fields weaker than $\sim 50$ mG can effectively couple to the fluid. At 5 AU, Ohmic diffusion dominates for $z/H \lesssim 1$ when the field is relatively weak ($B \lesssim$ a few milligauss), irrespective of the properties of the grain population. Conversely, at 10 AU this diffusion term is unimportant in all the scenarios studied here. High above the midplane ($z/H \gtrsim 5$), ambipolar diffusion is severe and prevents the field from coupling to the gas for all $B$. Hall diffusion is dominant for a wide range of field strengths at both radii when dust grains are present.

The growth rate, wavenumber and range of magnetic field strengths for which MRI-unstable modes exist are all drastically diminished when dust grains are present, particularly when they are small ($a \sim 0.1 \mu$m). In fact, MRI perturbations grow at $5$ AU (10 AU) for $B \lesssim 160$ mG (130 mG) when $3 \ \mu$m grains are mixed with the gas. This upper limit on the field 
strength is reduced to only $\sim 16$ mG (10 mG) when 
the grain size is reduced to $0.1 \ \mu$m. In contrast, when the grains are assumed to have settled, MRI unstable 
modes are found for $B \lesssim 800$ mG at 5 AU and 250 mG at 10 AU (Salmeron \& Wardle 2005). Similarly, as the typical size of the dust grains diminishes, the vertical extent of the dead zone increases, as expected. For $0.1 \ \mu$m grains, the disk is magnetically inactive within 
two scaleheights of the midplane at both radii, but perturbations grow over the entire section of the disk for grain sizes of $1 \ \mu$m or larger. When dust grains are mixed with the gas, perturbations that incorporate Hall diffusion grow faster, and are active over a more extended cross section of the disc,  than those obtained under the ambipolar diffusion approximation.

We conclude that in protoplanetary discs, the magnetic field is able to couple to the gas and shear over a wide range of fluid conditions even when small dust grains are well mixed with the gas. Despite the low magnetic coupling, MRI modes grow for an extended range of magnetic field strengths and Hall diffusion largely determines the properties of the perturbations in the inner regions of the disc.

\end{abstract}

\section{Introduction}
	\label{sec4:intro}
Magnetic fields may regulate the `disc accretion phase' of star 
formation by providing means of transporting away the excess angular momentum of the 
disc, enabling matter to accrete. The more generally relevant mechanisms associated 
with this are MHD turbulence induced by the magnetorotational instability (MRI; Balbus \& Hawley 1991, 1998) and 
outflows driven centrifugally from the disc surfaces (Blandford \& Payne 1982, Wardle \& 
K\"onigl 1993; see also the review by K\"onigl \& Pudritz 2000). These processes are, in turn, thought to play key roles in the dynamics and evolution of 
astrophysical accretion discs. Magnetically driven turbulence is likely to impact disc chemistry  (e.g. Semenov, Wiebe \& 
Henning 2006, Ilgner \& Nelson 2006) as well as the properties -- and evolution -- of dust 
grains mixed with the gas (e.g. Turner et al. 2006). Magnetic resonances have been shown to modify the net tidal torque exerted by the disc on 
a forming planet and thus, alter the speed and direction of the planet's migration through 
the disc (Terquem 2003; Johnson, Goodman \& Menou 2006). Finally, the magnetically inactive (dead) zones (Gammie 1996; Wardle 1997)  are not only the regions where planets are thought to form, but they have also been invoked  as a possible mechanism to stop their inward migration (e.g. Matsumura \& Pudritz 2005). 

In protoplanetary discs, however, the magnetic diffusivity can be high enough to limit -- or 
even suppress -- these processes. The specific role magnetic fields are able to play in 
these environments is, therefore, largely determined by the degree of coupling 
between the field 
and the neutral gas. A critical parameter for this analysis is the ionisation fraction of the fluid, which reflects the equilibrium between 
ionisation and recombination processes taking place in the disc. Ionisation processes 
outside the disc innermost sections 
($R \gtrsim 0.1$ AU) are non-thermal, driven by
interstellar cosmic rays, X-rays emitted by the magnetically active protostar and 
radioactive decay (Hayashi 1981; Glassgold, Najita \& Igea
1997; Igea \& Glassgold 1999; Fromang, Terquem \& Balbus 2002). On the other 
hand, free electrons are lost through recombination processes which, in general, take place both in the gas phase and on grain surfaces (e.g. Nishi, 
Nakano \& Umebayashi 1991). 

Dust grains affect the level of magnetic coupling in protoplanetary discs when 
they are well mixed with the gas (e.g. in relatively early stages of accretion
and/or 
when turbulence prevents them from settling towards the midplane). They do 
so in 
two ways. First, they reduce the ionisation fraction by providing additional 
pathways for electrons and ions to recombine on their surfaces. Second, 
charged dust particles can become important species in high density 
regions (Umebayashi \& Nakano 1990; Nishi, Nakano \& Umebayashi 1991). 
For example, at $1$ AU in 
a disc where $0.1 \ \mu$m grains are present, positively charged particles are the 
most abundant ionised species within two scaleheights from the midplane 
(see Fig.~6 of Wardle 2007; hereafter W07). As these particles generally have 
large cross sections, collisions with the neutrals are important and they 
become decoupled (or partially decoupled) to the magnetic field at densities 
for which smaller species, typically ions and electrons, would still be well 
tied to it. 

Both of these mechanisms act to lower the conductivity of the fluid, 
especially near 
the midplane where the density is high and ionisation processes are 
inefficient. In the minimum-mass solar nebula disc (Hayashi 1981, Hayashi et el. 1985), for example, X-rays are 
completely attenuated below $z/H \sim 1.7$ (see Fig.~$1$ of Salmeron \& 
Wardle 2005; hereafter SW05). As a 
result, in the disc inner sections the magnetic coupling may be
insufficient to provide adequate link between the neutral particles and the field. 
Moreover, recent calculations for a minimum-mass solar nebula disk exposed 
to X-ray and cosmic ray ionising fluxes (W07) indicate that near the surface (above 3 - 4 
tidal scaleheights) the magnetic diffusivity can also be severe, even though in these regions 
the ionising flux is strongest and the electron fraction is significantly larger than it is in the 
disc interior. This effect results from the strong decline in the number density of charged 
particles high above the midplane. These considerations suggest that magnetic activity in the inner regions of weakly ionised 
discs may well be confined to intermediate heights above the midplane ($z/H \sim 2 - 4
$). 

It is clear that the level of magnetic diffusion is strongly dependent on the presence, and size 
distribution, of dust particles suspended in the gas phase. In fact, once dust grains have settled, the ionisation fraction may be enough to produce adequate magnetic coupling over the 
entire vertical extension of the disk even at $R \lesssim 1$ AU (W07). A realistic study of the properties of the MRI in these discs 
must, therefore, incorporate a consistent treatment of dust dynamics and evolution 
(unless they are 
assumed to have settled, a good approximation to model relatively late 
accretion stages, as was the case in SW03 and SW05). This analysis is 
further complicated because dust grains have complex spatial and  
size distributions (e.g. Mathis, Rumpl \& Nordsieck 1977, 
Umebayashi \& Nakano 1990, D'Alessio et al. 2006), determined by the 
competing action of processes involving sticking, shattering, 
coagulation, growth (and/or sublimation) of ice mantles and settling to the 
midplane (e.g. Weidenschilling \& Cuzzi 1993). Previous results (SW03, SW05) 
highlight also the 
importance of incorporating in these studies all three diffusion mechanisms between the magnetic field and the neutral fluid (namely, the Ohmic, Hall and Ambipolar diffusivities), as 
Hall diffusion largely determines the 
growth and structure of MRI perturbations, particularly in the disc inner regions (e.g. within distances of the 
order of a few AU from the central protostar; SW05). This is implemented here via a diffusivity tensor formalism (Cowling 1957,
Norman \& Heyvaertz 1985, Nakano \& Umebayashi 1986, Wardle \& Ng 1999, W07). 

Dust grains can affect the structure and dynamics of 
accretion discs via two additional mechanisms: Dust opacity can modify the 
radiative transfer within the disc -- which, in turn, can dramatically alter 
its structure -- and dust particles may become dynamically important if their 
abundance is sufficiently high. In this study both effects are small because
the disc is vertically isothermal and grains constitute only a small fraction of the 
mass of the gas (see below).

In this paper we study the vertical structure
and linear growth of 
the MRI in a disc where dust grains are well mixed with the gas over its 
entire vertical dimension. Results are presented for two representative 
distances ($R = 5$ and $10$ AU) from the central protostar. For simplicity, we assume 
that
all particles have the same radius -- $a = 0.1$, $1$ or $3 \ \mu$m --  
and constitute $1$\% of the 
total mass of the gas, a typical assumption in studies of molecular 
clouds (Umebayashi \& Nakano 1990). This fraction is constant with height, 
which means that we 
have also assumed that no sedimentation has occurred. Although this is a very 
simplified picture, the results illustrate the importance of dust 
particles in the delicate ionisation equilibrium of discs, and consequently, 
on their magnetic activity. 

The paper is organised as follows. The 
adopted disc model is described in 
section \ref{sec4:cond}. Section \ref{sec4:formulation} briefly
summarises the formulation and methodology, which are based on SW05. We refer the 
reader to that study -- and references therein -- for further details. This section also 
includes a discussion of the
typical dependence of the components of the diffusivity tensor and 
magnetic coupling with height with (and without) grains for the two radial positions of 
interest. 
Section \ref{sec4:mri} then
presents the vertical structure and linear growth of unstable MRI modes at these radii and 
compares solutions incorporating different diffusion mechanisms and assumptions 
regarding the presence, and size, of dust grains. These results, and possible 
implications for 
the dynamics and evolution of low conductivity discs, are discussed in 
section \ref{sec4:discussion}. Our main conclusions are summarised in section \ref{sec:summary}.

\section{Disc model}
	\label{sec4:cond}

Our fiducial disc, assumed to be geometrically thin and vertically isothermal, is based on 
the minimum-mass solar nebula model (Hayashi 1981, Hayashi et al 
1985). Our formulation incorporates the disc 
vertical stratification but -- following common practice -- it neglects radial gradients. This is appropriate, as 
these gradients typically occur over a much larger length scale than those in the vertical direction. 
Under these 
assumptions, the equilibrium structure of the disc is the result of the 
balance between the vertical component of the gravitational force exerted by 
the central object and the pressure gradient within the disc. The vertical 
profile of the density is then given by 

\begin{equation}
\frac{\rho (r,z)}{\rho_{\rm 0}(r)}= \exp \left[-\frac{z^2}{2 H^2(r)}\right] \,,
	\label{eq:rhoinitial}
\end{equation}

\noindent
where $\rho_{\rm 0}$ is the midplane density and $H \equiv c_{\rm s}/\Omega$ is the tidal scaleheight of the gas. The neutral gas 
is assumed to be composed of molecular hydrogen and helium, such that 
$n_{\rm He} = 0.2 n_{\rm H_2}$, which results in $n_{\rm H}(r,z) = \rho(r,z)/1.4 m_{\rm 
H}$.

As already mentioned above, a key feature of protoplanetary accretion discs is that they 
are weakly ionised. This is because ionisation processes are generally ineffective 
(except possibly in the vicinity of the star and in the surface regions), 
while the recombination rate is accelerated 
by the high density of the fluid and the removal of charges by dust grains (if present). 
In fact, outside the innermost $0.1$ - $0.3
$ AU, where thermal effects are relevant (e.g. Hayashi 1981), the main ionising sources 
are X-rays and UV radiation emanating from the central object (e.g. Glassgold, Feigelson 
\& Montmerle 2000) and -- to a much lesser extent -- the decay 
of radioactive materials, particularly $\ ^{40} {\rm K}$ (Consolmagno \& Jokipii 1978, Sano 
et al. 2000). Interstellar cosmic rays may also be important as they can potentially reach deeper into the disc than X-ray and UV fluxes do. In fact, cosmic rays are the dominant ionising source at 1 AU for $z/H \lesssim 2.2$ (they even reach the midplane at this radius, albeit significantly attenuated; e.g. SW05). However, their actual contribution is unclear because the low-energy particles responsible for ionisation may be scattered by outflows launched from the protostar-disc system (e.g. Fromang et al. 2002). On the other hand, recombination processes in the disc generally occur both in the 
gas phase (through the 
dissociative recombination of electrons with molecular ions and the 
radiative recombination with metal ions) and on grain surfaces (e.g. Oppenheimer \& 
Dalgarno 1974, Spitzer 1978, Umebayashi \& Nakano 1980, Nishi et al. 1991, Sano et al. 
2000). For a typical abundance of metal atoms in the gas phase of $8.4 \ee{-5} \delta_{\rm 2}$\footnote{Here $\delta_{\rm 2} \approx 0.02$ is the fraction of heavy metal atoms in the gas phase, estimated from interstellar absorption lines in diffuse clouds (Morton 1974).}  (Umebayashi \& Nakano 1990), the radiative recombination rate of metal ions is dominant for all vertical locations of interest, with the exception of the uppermost sections of the disc (see Fig.~2 of SW05). 

The resulting ionisation fraction of the fluid largely determines the ability of the magnetic 
field to couple to the gas and shear and thus, regulates the magnetic activity in 
these astrophysical systems. 
In protoplanetary discs, in particular, the field has been envisioned to be dynamically 
important near the surface, whereas magnetic activity may 
be suppressed in their inner sections (the `layered accretion' scenario; Gammie 1996, 
Wardle 1997). However, the existence and configuration of a magnetically inactive -- 
dead -- zone in the disk interior has been shown to be critically dependent on the 
presence and properties of dust grains mixed with the gas (W07). As this study shows,  
in a minimum-mass solar nebula model at 1 AU the entire cross section of the disc is magnetically coupled 
when dust grains are assumed to have settled to the midplane. It is, in particular, the 
presence of small grains what most severely affects the magnetic coupling. For example, 
the presence of a standard interstellar population of $0.1 \ \mu$m grains at 1 AU reduces the total 
active layer of the disc from $\sim 1700 \  {\rm g} \ {\rm cm}^{-2}$ to $\sim 2 \  {\rm g} \ 
{\rm cm}^{-2}$. This column density increases to $\sim 80 \  {\rm g} \ {\rm cm}^{-2}$ once 
the grains aggregate to $3 \ \mu$m. At 5 AU, in contrast, the entire cross section of the 
disc is coupled once the grains have grown to $1 \ \mu$m (we refer the reader to W07 for further details of these models).

\section{Formulation and methodology}
        \label{sec4:formulation}

The solutions presented in this paper are based on the formulation detailed in SW03 and SW05. Only a brief summary is given here. We write
the equations of non-ideal MHD about a local Keplerian frame that corotates
with the disc at the Keplerian frequency $\Omega$ and express the velocity
field as a departure from exact Keplerian motion. We further assume that the 
abundances of charged species are sufficiently low to be able to neglect their inertia,  thermal 
pressure and the effect of ionisation and recombination processes on 
the neutrals. Under these conditions, only the equations
of motion for the neutral gas are required:

 \begin{equation}
\delt{\rho} + \div(\rho \v) = 0 \,,	
	\label{eq:continuity}
\end{equation}

\begin{equation}
\delt{\v} + (\v \cdot \grad)\v -2\Omega \vf \rh + \half\Omega \vr\fh 
-\frac{\vk^2}{r}\rh +
\frac{c_s^2}{\rho}\grad\rho +\grad \Phi = \frac{\J\cross\B}{c\rho}\,,
	\label{eq:momentum}
\end{equation}

\begin{equation}
\delt{\B} = \curl (\v \cross \B) - c \curl \E' 
-\thalf \Omega \B_r \fh \,.
	\label{eq:induction}
\end{equation}

\noindent
In the equation of motion (\ref{eq:momentum}), $\Phi$ is the 
gravitational 
potential for a non self-gravitating disc
and $\v_{\rm K}^2/r$ is the centripetal term generated by Keplerian 
motion. Coriolis terms $2\Omega \vf\rh$ and $\half \Omega 
\vr\fh$ are associated with the use of a local Keplerian 
frame and $c_{\rm s}$ is the isothermal sound speed.
In the induction equation (\ref{eq:induction}), $\E'$ is the electric field 
in the 
frame comoving with the neutrals and $\thalf \Omega \B_r \fh$ accounts for the
generation of toroidal field by the disc differential rotation. Finally, the magnetic 
field must also satisfy the constraint
$\div \B = 0$ and the current density must satisfy Ampere's law, 

\begin{equation}
\J = \frac{c}{4\pi}\grad\cross \B 
\label{eq1:j_curlB}
\end{equation}

\noindent
and Ohm's law 

\begin{equation}
	\J = \bmath{\sigma}\cdot \E' \,.
	\label{eq1:ohm}
\end{equation}

Following Wardle \& Ng (1999) and Wardle (1999; hereafter W99), the current density is 
expressed as,

\begin{equation}
	\J = \bmath{\sigma}\cdot \E' = \sigma_{\rm O} \Epa + 
	\sigma_{\rm H} \Bh \cross \Epe + \sigma_{\rm P} \Epe  \,,
	\label{eq:J-E}
\end{equation}

\noindent
where subscripts $\parallel$ and $\bot$ denote vector components parallel and 
perpendicular to $\mathbf{B}$. In this expression, $\sigma_{\rm O}$, $\sigma_{\rm H}$ 
and 
$\sigma_{\rm P}$ are the Ohmic, Hall and Pedersen conductivity terms %
given by,

\begin{equation}
	\sigma_{\rm O} = \frac{ec}{B}\sum_{j} n_j Z_j \beta_j \,,
	\label{eq1:sigma0}
\end{equation}

\begin{equation}
	\sigma_{\rm H} = \frac{ec}{B}\sum_{j}\frac{n_j Z_j}{1+\beta_j^2}
	\label{eq1:sigma1}
\end{equation}

\noindent
and

\begin{equation}
	\sigma_{\rm P} = \frac{ec}{B}\sum_{j}\frac{n_j Z_j \beta_j}{1+\beta_j^2} \,.
	\label{eq1:sigma2}
\end{equation}

\noindent
Subscript $j$ is used here to label the charged species. They are characterised by their 
number density $n_j$,
particle mass $m_j$, charge $Z_j e$ and Hall parameter

\begin{equation}
\beta_j= {Z_jeB \over m_j c} \, {1 \over \gamma_j \rho} 
\label{eq1:Hall_parameter}
\end{equation}

\noindent
(the ratio of the gyrofrequency and the collision frequency with the 
neutrals), which measures the relative importance of the Lorentz and drag forces in balancing 
the electric force on the particle. 

Equation (\ref{eq:J-E}) can be inverted to find an expression for $\E'$. This leads 
to the following form of the induction equation (W07)

\begin{eqnarray}
\lefteqn{\delt{\B} = \curl \left (\v \cross \B\right) - \curl \left[ \eta_{\rm O} \curl \B \right. }
	\nonumber\\
& & {} + \left.  \eta_{\rm H} (\curl \B) \cross \hat{\B} + \eta_{\rm A} (\curl \B)_{\perp} \right] 
\,,
	\label{eq1:induction_drift}
\end{eqnarray}

\noindent
where

\begin{equation}
\eta_{\rm O} = \frac{c^2}{4 \pi \sigma_{\rm O}} \,,
\label{eq:etaO}
\end{equation}

\begin{equation}
\eta_{\rm H} = \frac{c^2}{4 \pi \sigma_{\perp}} \frac{\sigma_{\rm H}}{\sigma_{\perp}} 
\label{eq:etaH}
\end{equation}

\noindent
and

\begin{equation}
\eta_{\rm A} = \frac{c^2}{4 \pi \sigma_{\perp}} \frac{\sigma_{\rm P}}{\sigma_{\perp}} - 
\eta_{\rm O} 
\label{eq:etaA}
\end{equation}

\noindent
are the Ohmic, Hall and ambipolar diffusivities; and 

\begin{equation}
\sigma_{\perp} = \sqrt{\sigma_{\rm H}^2 + \sigma_{\rm P}^2} 
\end{equation}

\noindent
is the total conductivity perpendicular to the magnetic field. When ions and electrons are the only charged species, it can be shown that (W07)

\begin{equation}
|\etaH| = |\beta_{\rm e}| \etaO
\label{eq:etaH1}
\end{equation}

\noindent
and 

\begin{equation}
\etaA =  |\beta_{\rm e}| \beta_{\rm i} \etaO \,.
\label{eq:etaA1}
\end{equation}

Note that the 
Ohmic ($\etaO$) and ambipolar ($\etaA$) diffusivity terms
are always positive, as the former does not depend on the magnetic field strength and the second 
scale quadratically with it. As a result, they are both invariant under a reversal 
of the magnetic field polarity. On the contrary, the Hall term ($\etaH$) scales linearly with 
$B$ and thus, can become negative. The change in sign of $\etaH$ corresponds, in turn,  
to a change in the 
direction of the magnetic field at the height where particular species become
decoupled to it by collisions with the neutrals. It corresponds, therefore, to changes in 
the contribution of different charged species to this component of the 
diffusivity tensor. 

The relative importance of the diffusion terms in (\ref{eq:etaO}) to (\ref{eq:etaA}) differentiate three \emph
{diffusivity regimes}:

\begin{enumerate}
\item In the \emph{Ambipolar diffusion} regime, $|\beta_j| \gg 1$ for most charged 
species and
$\etaA \gg |\etaH| \gg \etaO$. In this limit, which is typically dominant in low density regions (e.g. in molecular clouds and near the surface of protoplanetary discs), the magnetic field is effectively frozen into the 
ionized component of the fluid and drifts with it through the neutrals.
\item \emph{Ohmic (resistive)} limit. In this case $|\beta_j| \ll 1$ for most charged 
species, resulting in 
$\etaO \gg |\etaH| \gg \etaA$. The magnetic field can not be regarded as being frozen into 
any fluid component and the diffusivity is a scalar, the well-known Ohmic diffusivity. This regime dominates close to the midplane in the inner regions ($R \lesssim 5$ AU) of protoplanetary discs when the magnetic field is relatively weak (W07).
\item \emph{Hall diffusion} limit, which occurs when $|\beta_j| \gg 1$ for charged species 
of one sign (typically electrons) and $\ll 1$ for those of the other sign (e.g.~ions). In this 
case, $|\etaH| \gg \etaA$ and  $\etaO$. This regime is important at intermediate densities (between those at which ambipolar and Ohmic diffusion regimes dominate). It has been shown to prevail under fluid conditions satisfied over vast regions in protoplanetary discs (e.g. Sano \& Stone 2002a, W07).
\end{enumerate}

The diffusivities used in this study were calculated using the procedure described in W07, 
to which we refer the reader for details. Essentially, the adopted chemical reaction 
scheme is based on that of Nishi, Nakano \& Umebayashi (1991), but it allows for higher charge 
states on dust grains. This is necessary because of the higher temperature and density of protoplanetary discs in relation to those typically associated with molecular clouds.

Finally, in the following sections we will also use the magnetic coupling parameter (W99)

\begin{equation}
\chi \equiv \frac{\omega_{\rm c}}{\Omega} = \frac{1}{\Omega}\frac{B_{\rm 0} \sigma_{\perp}}{\rho c^2} \,,
\label{eq:chi} 
\end{equation}

\noindent
the ratio of the critical frequency ($\omega_{\rm c}$) \emph{above} which flux-freezing conditions break down
and the dynamical (Keplerian) frequency of the disc. For the sake of clarity we now sketch the arguments leading to this expression (see also Wardle \& Ng 1999). First, recall that non-ideal MHD effects become important when the inductive and diffusive terms in the induction equation (\ref{eq:induction}) are comparable; or

\begin{equation}
\curl (\v \cross \B) \sim c \curl \E' \,,
\label{eq:ind1}
\end{equation}

\noindent
where (see equation \ref{eq1:ohm})

\begin{equation}
\E' = \frac{\J}{\sigma} = \frac{1}{\sigma} \left( \frac{c}{4\pi}\grad\cross \B \right) \,.
\label{eq:ohm10}
\end{equation}

\noindent
In expression (\ref{eq:ohm10}), $\sigma$ is taken to be a characteristic measure of the conductivity of the gas. Next, we adopt the following typical values for the various terms above

\begin{displaymath}
\grad \sim k = \omega/v_{\rm A} \qquad
v \sim v_{\rm A} \qquad
B \sim B_{\rm 0} \qquad
\sigma \sim \sigma_{\perp} \,,
\end{displaymath}

\noindent
where $k$ is the wavenumber of the MRI perturbations in flux-freezing conditions and $v_{\rm A} \equiv B^2/4 \pi \rho$ is the local Alfv\'en speed. Substituting these relations in (\ref{eq:ind1}) yields the 
desired expression for $\omega_{\rm c}$. This parameter is useful because
in ideal-MHD conditions, the 
instability grows at $\sim 0.75 \Omega$ (Balbus \& Hawley 1991). Consequently, if $\omega_{\rm c} < \Omega$ (or $\chi 
< 1$) the field is poorly coupled to the disc at the frequencies of interest for the analysis of 
the MRI and non-ideal MHD effects are expected to be important.

Equations (\ref{eq:continuity}) to (\ref{eq1:j_curlB}) and (\ref{eq:J-E}) are linearized about 
an initial (labeled by a subscript `0'), steady state where the 
magnetic field is vertical and the current density, fluid velocity and electric
field in the frame comoving with the neutrals all vanish. As a result of the
last condition, the changes in the components of the diffusivity tensor are
not relevant in this linear formulation (e.g. $\E_{\rm 0}' \cdot \delta \bmath{\sigma} \equiv 0$) and only the initial, unperturbed 
values are required.   
Taking perturbations of the form 
$\mathbf{q} = \mathbf{q}_{0} + \mathbf{\delta q}(z) \e^{i\omega t}$ and 
assuming $k = k_z$ we obtain a system of ordinary 
differential equations (ODE) in $\dE$ (the 
perturbations of the electric field in the laboratory frame), $\dB$, and 
the perturbations' growth rate $\nu=i\omega/\Omega$. Three parameters are found to control
the dynamics and evolution of the fluid: 
($1$) $v_{\rm A}/c_{\rm s}$, the local ratio of the Alfv\'en 
speed and the isothermal sound speed of the gas, which measures the strength 
of the magnetic field. 
($2$) $\chi$, the coupling between the ionised and neutral components of the fluid.
($3$) $\eta_{\rm H}/\eta_{\rm P}$, 
the 
ratio of the diffusivity terms perpendicular to $\mathbf{B}$, 
which characterises the diffusivity regime of the fluid.
These parameters are evaluated at different locations ($r$, $z$) of the 
disc 
 taking
the magnetic field strength ($B > 1$ mG) as a free parameter. The 
system of equations is integrated vertically as a two-point boundary value 
problem for coupled ODE with boundary conditions 
$\delta B_r = \delta B_{\phi} = 0$ and $\delta E_r=1$ at $z=0$ 
and $\delta B_r = \delta B_{\phi} = 0$ at $z/H=6$. 

Given that magnetic diffusion can have such a dramatic effect on the properties of magnetically-driven turbulence in protoplanetary discs,  we now present calculations of the magnetic diffusivity at 
5 and 10 AU. We explore which disc regions are expected to be magnetically coupled 
and which diffusion mechanism is dominant at different positions ($r$, $z$) as a function of $B$. This discussion is relevant for the analysis of our MRI results at these locations (section \ref{sec4:mri}).

\subsection{Magnetic diffusivity}
	\label{subsec:diffusivity}
	
\begin{figure} 
\centering
 \includegraphics[width=0.45\textwidth]{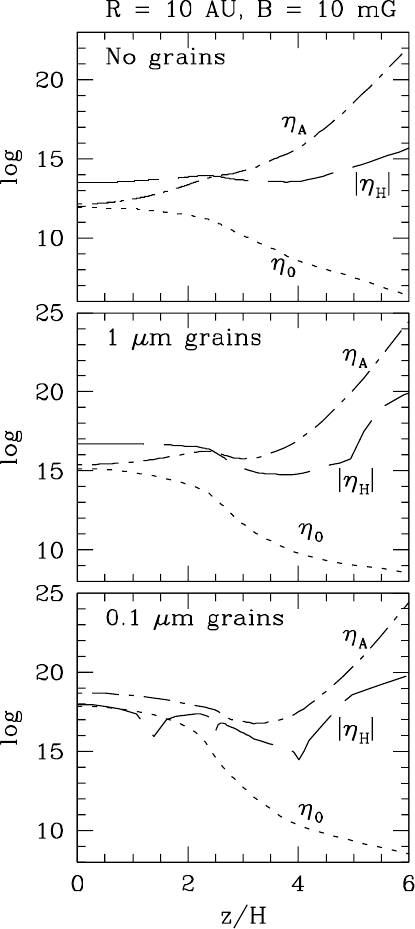} 
\caption{Components of the diffusivity tensor ($\eta_{\rm O}$,  $|\eta_{\rm H}|$ and $
\eta_{\rm A}$) as a function of height for $R = 10$ AU and $B = 10$ mG. In the top 
panel, dust grains have settled to the midplane. The middle and bottom panels show the solutions when dust grains of radius $a = 1$ and 0.1 $\mu$m, respectively, are well mixed with the gas over the entire disc thickness. When the grains are $\gtrsim 1 \ \mu$m in size (or are absent), Hall diffusion dominates for $z/H \lesssim 
2.5$. In contrast, for the 0.1 $\mu$m-sized grains, $\etaA > |\etaH|$ for all $z$ and $|\etaH| > \etaO
$ above $z/H \sim 1$. Note that $|\etaH|$ shows `spikes' at the $z$ where it changes 
sign in response to different charged species becoming decoupled to the magnetic field by collisions 
with the neutrals. }
\label{fig:10AU_eta_1}
\end{figure}

Fig.~\ref{fig:10AU_eta_1} shows the components of the diffusivity tensor 
($\etaO$, $|\etaH|$ and $\etaA$) 
as a function 
of height for $R = 10$ AU and a representative field strength ($B = 10$ mG). The solutions in the top panel have been obtained assuming that dust grains
have settled into a thin layer about
the midplane, so the charges are carried by ions and electrons only. The results in the middle and bottom panels incorporate a population of 1 and 0.1 $\mu$m-sized 
grains, respectively.
Note that when dust grains are present, all diffusivity terms 
increase drastically in relation to their values in the no-grains scenario. This 
effect becomes more accentuated as the grain size diminishes, given the 
efficiency of small grains in removing free electrons from the gas. For example, in the case that incorporates  
$0.1 \ \mu$m-sized particles (bottom panel), the diffusivity components at the midplane are larger, by 4 -- 7  
orders of magnitude, than their corresponding values when the grains have 
settled. 

When dust grains have either settled or aggregated to at least 1 $\mu$m in size 
(see top and middle panels), 
$|\etaH|$ is dominant for $z/H 
\lesssim 2.5$ and the fluid in this region is in the Hall diffusivity  regime. Also, $\etaA > \etaO$ there which implies that, if ions and electrons are the sole charge carriers, $|\beta_{\rm e}| \beta_{\rm i} > 1$ (see equation \ref{eq:etaA1}). For higher $z$, 
$\etaA > |\etaH| > \etaO$ and ambipolar diffusion 
dominates.  Note that the Hall diffusivity 
term increases less sharply than the ambipolar diffusivity in response to the 
fall in fluid density. This is a general feature, and an expected one, given that the former scales with $\rho^{-1}$ and the latter 
with $\rho^{-2}$. As a result, $\etaA$ is typically several orders 
of magnitude larger than $|\etaH|$ near 
the surface regions of the disc. 
Finally, for the $0.1 \ \mu$m-sized grain population (bottom panel), $\etaA > |\etaH|$ for all  $z$. This can be traced back to the fact that Hall diffusion is suppressed here because of the nearly equal abundances of (negatively charged) grains and ions in this region. 

Note also that the Hall diffusivity 
component shows characteristic `spikes', at the heights where it changes 
sign. This effect is also particularly evident for the $0.1 \ \mu$m-sized grains. In this 
scenario, in particular, $\etaH$ is positive when $0 \lesssim z/H \lesssim 1.6$
and $2.5 \lesssim z/H \lesssim 4.1$. It is negative at all other vertical 
locations. As mentioned above, this corresponds to different charged species contributing to 
this diffusion term. In order to explore the contributions to $\etaH$ in 
each of the vertical sections in which it has a different sign, we now describe 
the 
behaviour of the charged species at four representative heights, namely 
$z/H = 1$ and $3$ (for which $\etaH > 0$), and $z/H = 2$ and $5$ (where 
$\etaH < 0$). 

At $z/H = 1$, the density is sufficiently high for electrons to be able to stick to 
the grains. As a result, they reside mainly on dust particles, as do about a 
third of the ions. The contribution of (negatively charged) grains and ions 
to the Hall diffusivity term are very 
similar, with a small positive excess, which determines the sign of 
$\etaH$ in this region. On the contrary, at $z/H = 2$, ions are the 
dominant positively charged species, while the negative charges are still 
contained in dust grains and drift together with the neutrals. 
Consequently, $\etaH$ is negative in this section of the disc. 
At $z/H = 3$, ions and electrons are 
the main charge carriers and ions dominate the contribution to the Hall term,
which makes $\etaH$ positive. Finally, for $z/H = 5$, the dominant 
contribution to the Hall diffusivity term comes from the small percentage 
of remaining 
negatively charged grains. As a result, $\etaH$ is negative (and very 
small). 

\begin{figure} 
\centering
 \includegraphics[width=0.45\textwidth]{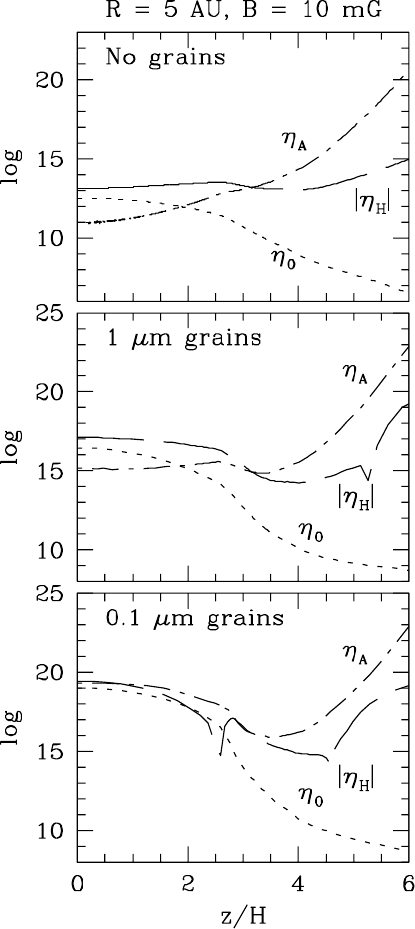} 
\caption{As per Fig.~\ref{fig:10AU_eta_1} for R = 5 AU. Note that when the grains are $\gtrsim 1 \mu$m in size or are absent (top and middle panels), Hall diffusion dominates for $z/H \lesssim 3$. At this radius, however, $\etaO > \etaA$ for $z/H \lesssim 2$.}
\label{fig:5AU_eta_1}
\end{figure}

We now turn our attention to $R = 5$ AU (Fig.~\ref{fig:5AU_eta_1}). At this radius, when the dust particles are at least $1\  \mu
$m in size (or are absent), Hall diffusion dominates for $z/H \lesssim 3$. In contrast with 
the solutions at 10 AU, however, Ohmic diffusion dominates over ambipolar diffusion ($\etaO > \etaA$) for most of this region (which implies that $|\beta_{\rm e}| \beta_{\rm i} < 1$ in this section of the disc when no grains are present). This is expected, given the larger fluid density at this radius in comparison with the 10 AU case discussed above. When the grains are small ($a = 0.1 \ \mu$m, see bottom panel), ambipolar diffusion is dominant ($\etaA > |\etaH| > \etaO$) for $1 \lesssim z/H \lesssim 2.8$ and $z/H \gtrsim 3.5$. At all other heights, $\etaA \approx |\etaH| > \etaO$. 

\begin{figure} 
\centering
 \includegraphics[width=0.4\textwidth]{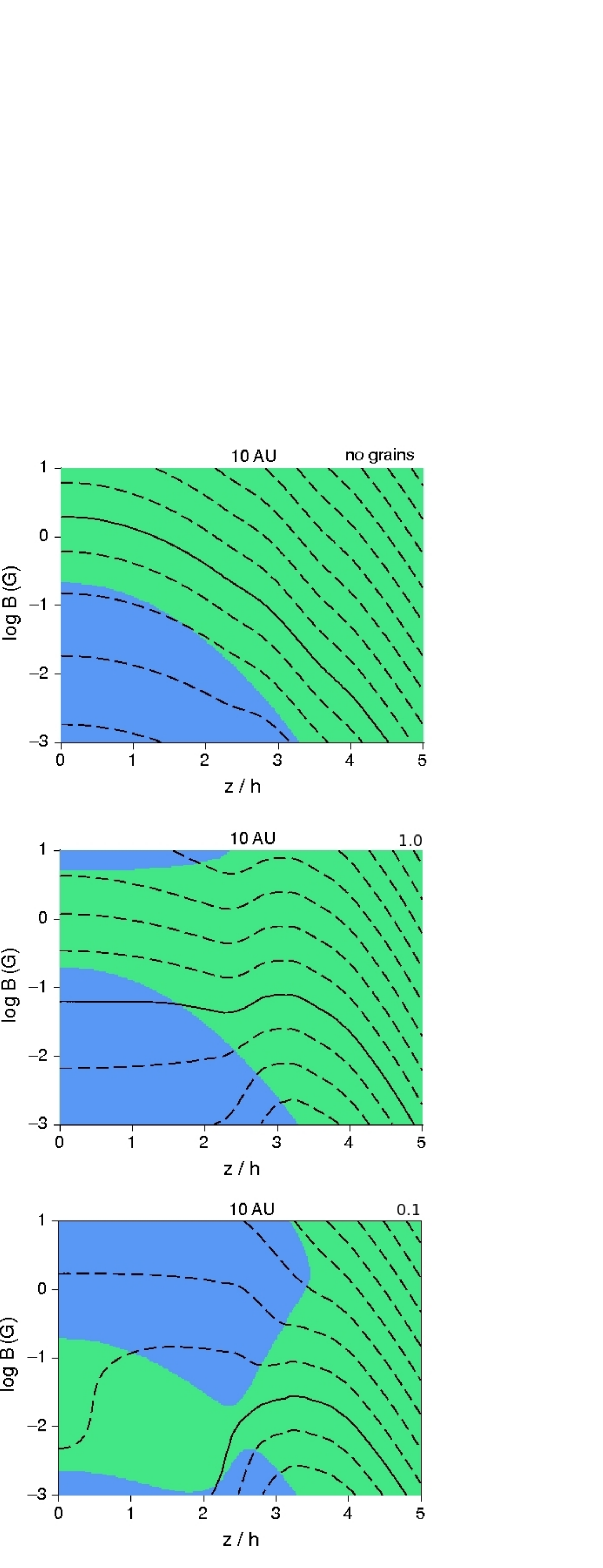} 
\caption{Contours of $\tilde{\eta} \equiv (\etaO^2 + \etaH^2 + \etaA^2)^{1/2}$ in a log(B) - $z/
H$ plane for $R = 10$ AU.  In the top panel, dust particles are assumed to have settled to the midplane. The effect of dust grains is included in the middle ($a = 1 \ \mu$m) and bottom ($a = 0.1 \ \mu$m) panels. The solid line is the critical value of the 
diffusivity $\tilde{\eta}_{\rm crit} \equiv H c_{\rm s}$ (W07) above which the magnetic field does not 
couple to the gas and shear. The background shading (from dark to light) denotes the 
dominant diffusion mechanism as Ohmic, Hall or Ambipolar. Note that when the grain size is 1 $\mu$m or larger (or they have settled), the disc midplane is magnetically coupled for a range of $B$. In contrast, when the grains are small (bottom panel), magnetic diffusion prevents the field to couple to the gas for $z/H \lesssim 2$. }
\label{fig:contour_10AU}
\end{figure}

\begin{figure} 
\centering
 \includegraphics[width=0.4\textwidth]{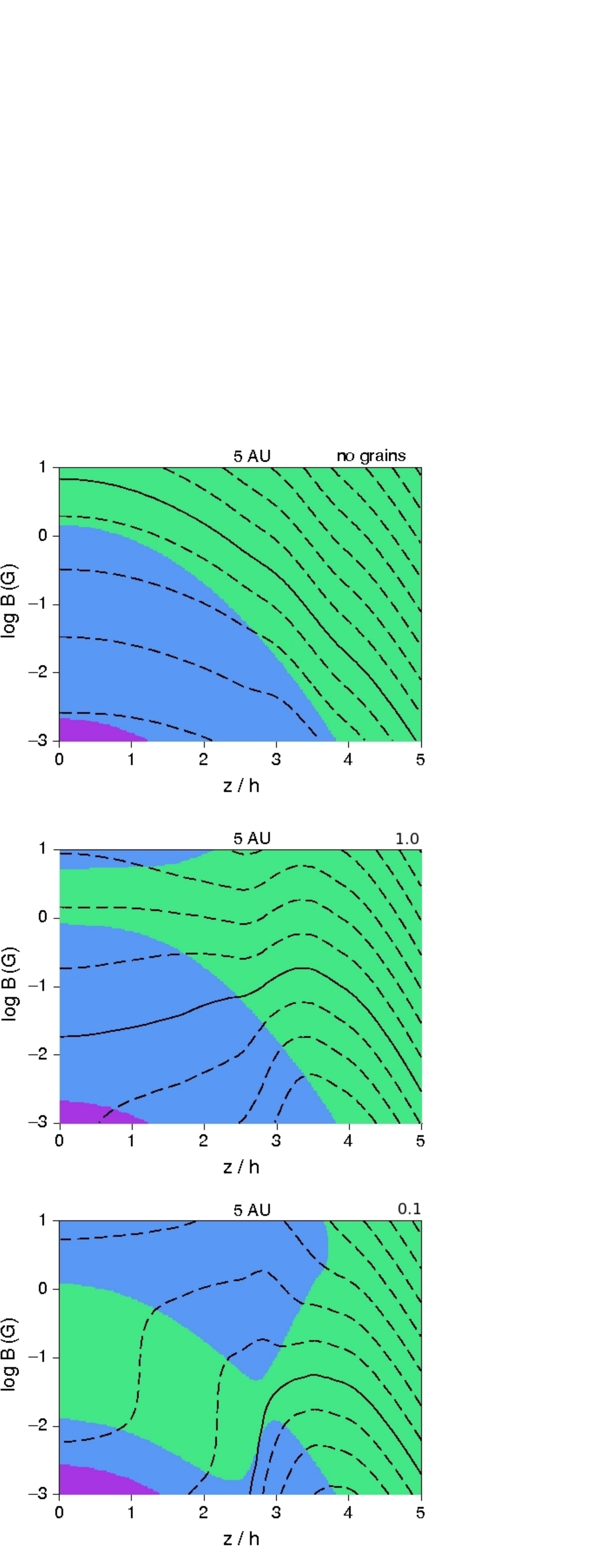} 
\caption{As per Fig.~\ref{fig:contour_10AU} for $R = 5$ AU. Note that in all cases, Ohmic diffusion is dominant close to the midplane ($z/H \lesssim 1.2$) when the field is weak ($B < $ a few mG). The midplane can be magnetically coupled when the grains are absent or have aggregated to $a \gtrsim 1 \ \mu$m. However, a dead zone develops below $\sim 2.5$ scaleheights when they are small ($a = 0.1 \ \mu$m). In this scenario, only $B \lesssim 50$ mG can couple to the gas (and only for $z/H \gtrsim 2.5$). Note that ambipolar diffusion dominates for all $B$ when $z/H \gtrsim 4$ (all panels) and for 1 mG $\lesssim B \lesssim 10 - 100$ mG below this height (bottom panel only).}
\label{fig:contour_5AU}
\end{figure}

Figs.~\ref{fig:contour_10AU} and \ref{fig:contour_5AU} generalise the analysis of the previous paragraphs to 
other field strengths for $R = 10$ and $5$ AU, respectively. As before, the top panels refer to discs where no grains are present. The middle and bottom panels incorporate 1 and 0.1 $\mu$m-sized grains, respectively. In these plots, the contours show the 
values of $\tilde{\eta} \equiv (\etaO^2 + \etaH^2 + \etaA^2)^{1/2}$ and the background 
shading (from dark to light) denotes the dominant diffusion mechanism as Ohmic, Hall or 
Ambipolar. The solid line is the critical value of the diffusivity $\tilde{\eta}_{\rm crit} \equiv H c_{\rm s}$ 
(W07), above which the diffusion term in the induction equation $|\curl (\tilde{\eta} 
\curl \B)|$  is larger than the inductive term $|\curl (\v \cross \B)|$  (see equation \ref
{eq:induction}), a situation that effectively limits the ability of the field to couple to the 
Keplerian shear. The dashed lines correspond to increases (decreases) in $\tilde{\eta}$ 
by factors of 10 in the direction of a stronger (weaker) magnetic field. 

Note that at 10 AU, in the no-grains case (top panel of Fig.~\ref{fig:contour_10AU}), field strengths $\lesssim$ a few Gauss are able to couple to the gas at the midplane. Hall diffusion dominates at this location for $B \lesssim 0.1$ G but the range of field strengths over which this occurs decreases gradually with height, so that when $z/H \gtrsim 3.3$ ambipolar diffusion is dominant for all $B$. The coupled region potentially extends to $z/H \sim 4.5$, depending on the field strength. When the particles have 
aggregated to 1 $\mu$m in size (middle panel of Fig.~\ref{fig:contour_10AU}), the situation is qualitatively similar to the one just discussed, the only significant difference being that in this case Hall diffusion is dominant at the midplane for all $B$ that can couple to the gas.

The previous results, however, are significantly modified when small dust grains are present
($a = 0.1 \ \mu$m, bottom panel). In this case, the section of the disc below two scaleheights 
is magnetically inactive and the magnetic diffusivity is severe enough to prevent coupling 
over the entire disc thickness when $B \gtrsim 25$ mG.  Ambipolar 
diffusion dominates for all field strengths above $\sim$ four scaleheights and also for $\sim 1$ mG $\lesssim B \lesssim 10 - 100$ mG below this height (the actual upper limit of this 
range varies with $z$). Hall diffusion is inhibited close to the midplane for this range of field strengths
because the number density of positively and negatively charged grains are very similar. 
Note, finally, that Ohmic diffusion is not dominant in any of the depicted scenarios at this radius, as expected, given the relatively low density of the fluid.

The results at 5 AU are qualitatively similar (Fig.~\ref{fig:contour_5AU}) to the ones just discussed. One key difference, however, is that Ohmic diffusion dominates in all cases for 
$z/H \lesssim 1.2$ provided that the field is sufficiently weak ($B \lesssim
$ a few milligauss). Ambipolar diffusion, on the other hand, is dominant for relatively strong fields (e.g. $B \gtrsim 1$ G at $z = 0$ in the no-grains case) as well as for all $B$ near the disk surface ($z/H \gtrsim 4$). Hall diffusivity is the most 
important diffusion mechanism for fluid conditions in-between those specified above. 
Note that the coupled region extends up to $z/H \sim 5$ in all cases. When no grains are present (top panel) the midplane is coupled for fields up to a few Gauss. This upper limit drops to $\sim 20$ mG when $1 \ \mu$m grains are mixed with the gas. When the grains are $0.1 \ \mu$m in size, however, only the 
region above $z/H \sim 2.5$ is coupled, and only for $B \lesssim 50$ mG. This `dead' zone is slightly more extended at this radius than at 10 AU, but stronger fields are able to couple to the gas in this case. The steep 
contours close to the midplane in this scenario result from the pronounced increase in the diffusivity in response to the increase in fluid density and the decline in the ionisation 
fraction of the gas as $z$ diminishes. 

\subsection{Magnetic coupling and MRI unstable modes}
\label{subsec:coupling}

In this section we address the following question: Which magnetic diffusion mechanism determines the properties of the MRI in different 
regions in protoplanetary discs? In this connection, it is useful to recall (see W99, SW03 and references therein for details) that in the disc regions where 
the magnetic coupling $\chi > 10$, ideal-MHD conditions 
hold and the particular configuration 
of the diffusivity tensor has little effect on the behaviour of the MRI. 
When $\chi$ is weaker than this but $\gtrsim |\sigma_{\rm H}|/\sigma_{\perp}$, 
ambipolar diffusion dominates. Finally, in 
the regions where $\chi < |\sigma_{\rm H}|/\sigma_{\perp}$, Hall 
diffusion 
modifies the structure and growth of MRI unstable modes, provided that this degree of
coupling is sufficient for unstable modes to grow. 

\begin{figure} 
\centering
 \includegraphics[width=0.4\textwidth]{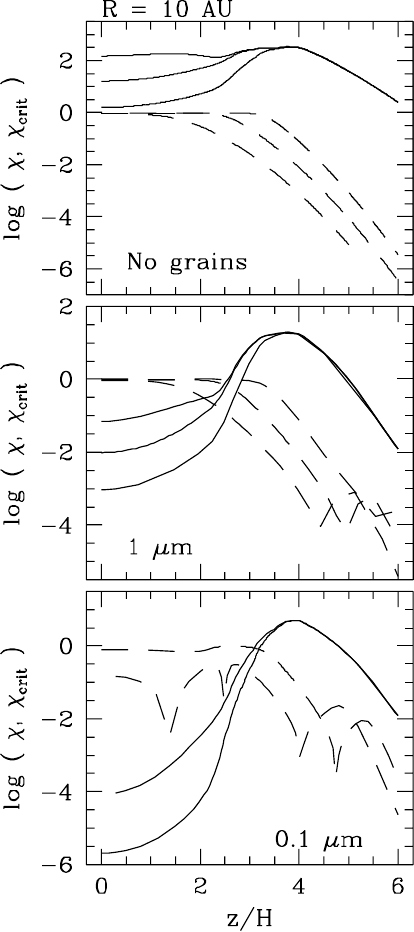} 
\caption{Comparison of the local magnetic coupling $\chi$ (solid lines) and critical 
coupling $\chi_{\rm crit} \equiv |\sigma_{\rm H}|/\sigma_{\perp}$ (dashed lines) for $R = 
10$ AU as a function of the magnetic field strength and for different assumptions regarding the presence (and size) of dust grains. Hall diffusion modifies the structure and growth rate of MRI unstable modes in the regions where $\chi < \chi_{\rm crit}$ provided that the coupling is sufficient for the instability to operate (SW03). Ambipolar 
diffusion is important if $\chi_{\rm crit} < \chi \lesssim 10$. For stronger $\chi$, ideal-MHD 
describes the fluid adequately (W99). The bottom (and leftmost) lines correspond to $B 
= 1$ mG in all panels. $B$ increases by a factor of 10 between curves (towards larger $
\chi$), except that the top curve for $a = 1 \ \mu$m grains corresponds to $B = 77$ mG, the 
strongest field for which perturbations grow. }
\label{fig:chi_10AU}
\end{figure}

\begin{figure} 
\centering
 \includegraphics[width=0.4\textwidth]{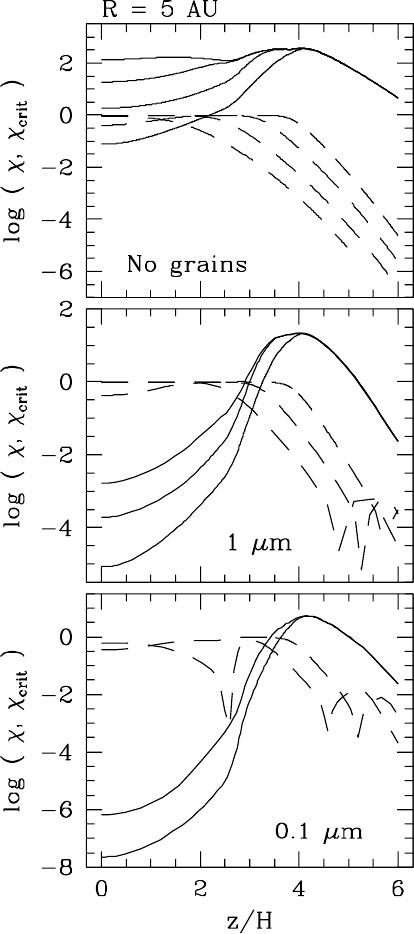} 
\caption{As per Fig.~\ref{fig:chi_10AU} for $R = 5$ AU. Note, however, that the top (and rightmost) lines 
correspond to $B = 795$ mG (top panel) and $B = 83$ mG (middle panel), the maximum 
field strength for which unstable modes grow in each scenario.}
\label{fig:chi_5AU}
\end{figure}

Figs.~\ref{fig:chi_10AU} and \ref{fig:chi_5AU} compare $\chi$ with the ratio 
$|\sigma_{\rm H}|/\sigma_{\perp}$ at $R = 10$ and $5$ AU, respectively, as a function of 
the magnetic field strength and for different assumptions regarding the presence, and 
radius, of dust grains mixed with the gas. In each figure, the top panels depict the results obtained assuming that no grains are present whereas the effect of 1 $\mu$m (0.1 $\mu$m)-sized grains is shown in the middle (bottom) panels. Note the dramatic impact dust grains have in the level of 
magnetic coupling at both radii. For example, introducing a population of $0.1 \ \mu$m grains causes $\chi$ to drop by 6 - 8 orders of magnitude at the midplane. 
Bottom (solid) and leftmost 
(dashed) lines correspond to $B = 1$ mG in all cases. The field strength increases by a 
factor of ten towards larger $\chi$, except that at both radii the top (and rightmost) curves for $a = 1 \ \mu$m show the maximum field strength for which the MRI grows. This is also the case for the solutions with no grains at 5 AU (the maximum values of $B$ for these cases are noted in the captions of Figs.~\ref{fig:chi_10AU} and \ref{fig:chi_5AU}).

Note that at $10$ AU (Fig.~\ref{fig:chi_10AU}), Hall diffusion is not expected to play 
an important role in the local properties of the MRI once the grains have settled ($\chi > |\sigma_{\rm H}|/\sigma_{\perp}$ for all $z$ and $B$). 
In this scenario, ambipolar diffusion dominates in the inner sections of the disc when the field is weak ($z/H \lesssim 2$ and $B \lesssim 10$ mG) while ideal-MHD holds at all  heights for stronger $B$.  On 
the contrary, when either 1 or 0.1 $\mu$m grains are mixed with the gas (middle and bottom panels),  
Hall diffusion has an impact on the MRI within $\sim$ three scaleheights of the midplane. 
At higher $z$, the ionisation fraction is such
that $|\sigma_{\rm H}|/\sigma_{\perp} \lesssim \chi \lesssim 10$ and ambipolar 
diffusion determines the local properties of MRI unstable modes.

The corresponding solutions at $5$ AU are shown in Fig.~\ref{fig:chi_5AU}. In this case, Hall diffusion dominates within two scaleheights from the midplane if the magnetic field is weak ($B  \lesssim 10$ mG) and the grains have settled (top panel). This is consistent with the higher 
column density at this radius in comparison to the 10 AU case discussed above (for which Hall diffusion was unimportant). Ambipolar diffusion is dominant in 
this scenario for $10$ mG $\lesssim B \lesssim 100$ mG but for stronger 
fields, $\chi > 10$ for all $z$ and the fluid is in ideal-MHD conditions over the entire 
section of the disc. On the other hand, when dust grains are present (middle and bottom panels), Hall diffusion determines the properties of MRI-unstable modes in the inner sections of 
the disc ($z/H \lesssim 3$) for all magnetic field strengths for which they grow. Ambipolar diffusion is locally dominant at higher $z$. 

Finally, note that when dust grains are present, the extent of the disc section where Hall diffusion has an impact on the MRI at both radii is quite insensitive to the strength of the field. Evidently,  dust particles can efficiently reduce the degree of magnetic coupling in the disc inner sections for a wide range of magnetic field strengths.
In light of the concepts presented so far, we now analyse the properties of the MRI at the two radii of interest.

\section{Magnetorotational Instability}
	\label{sec4:mri}

\begin{figure*} 
\centering
 \includegraphics[width=0.8\textwidth]{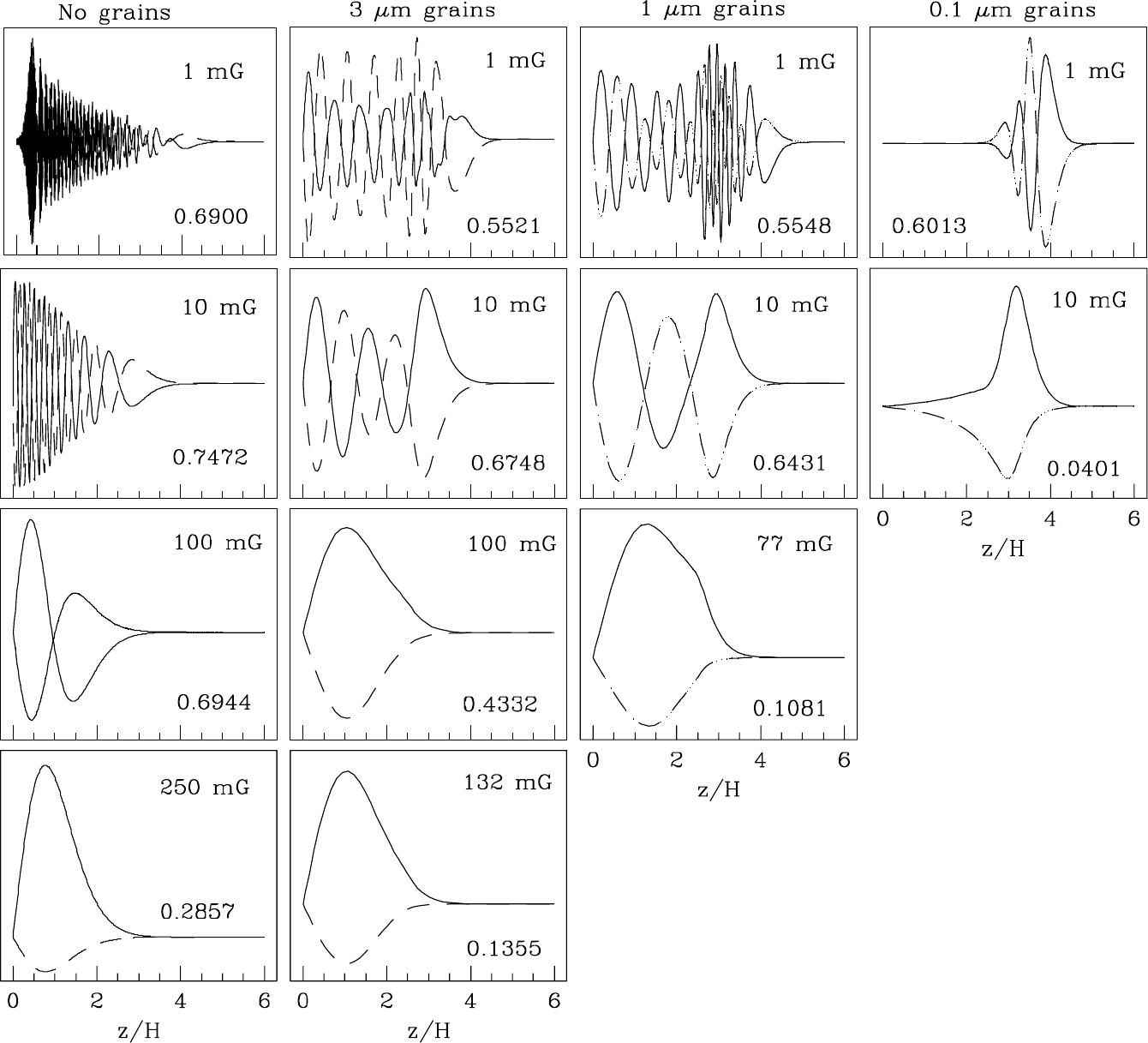} 
\caption{Structure and growth rate of the fastest growing MRI modes for 
$R = 10$ AU and different choices of the magnetic field strength. The leftmost
column shows the perturbations obtained if dust grains have settled, while the remaining ones, from 
left to right, 
display results assuming a population of single-sized grains of radius $a = 3$, 1 and 0.1 $\mu$m, 
respectively, are well mixed with 
the gas. The growth rate is indicated in the lower right corner of 
each panel. The strength of the field appears in the top right corner. Results are 
displayed for $B$ spanning from 1 mG, the weakest magnetic field for which unstable 
modes could be computed, to the maximum strength for which unstable modes were 
found in each case. Note 
the reduced wavenumber, growth rate and range of $B$ for which perturbations exist -- as 
well as the extended dead zone -- when dust grains are present. When the grains are relatively large (central two columns) Hall diffusion controls the modes, which grow even at the midplane. In contrast, when they are small (right column), the magnetic coupling is too low to sustain the instability for $z/H \lesssim 3$ and ambipolar diffusion controls the perturbations that grow above this height.}
\label{fig:grains1}
\end{figure*}

\begin{figure*} 
\centering
 \includegraphics[width=0.8\textwidth]{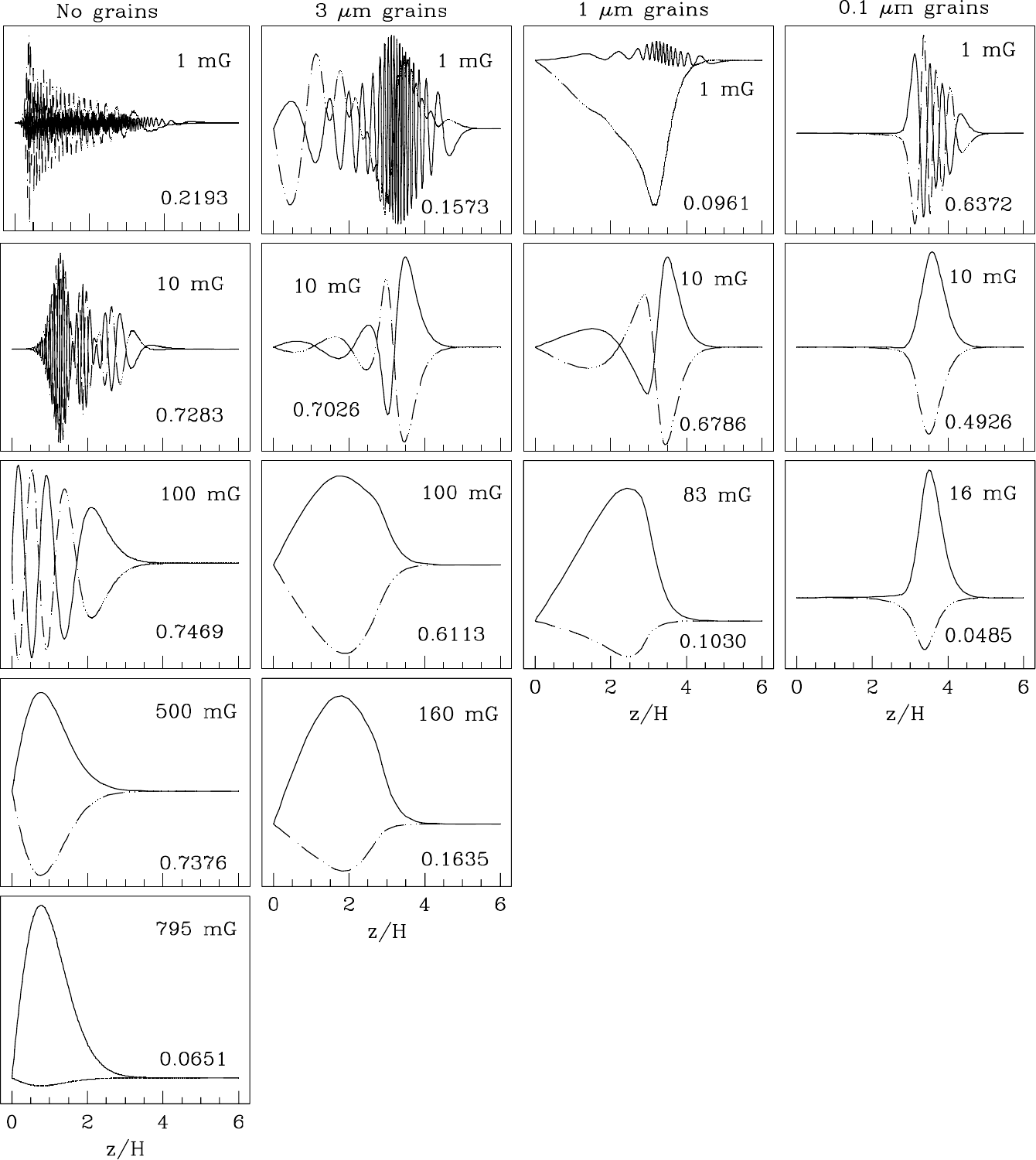} 
\caption{As per Fig.~\ref{fig:grains1} for $R = 5$ AU. At this radius, the solutions that incorporate 3 and 1 $\mu$m-sized grains (central two columns) are shaped by ambipolar diffusion, as this mechanism is dominant at the height where they peak (see middle panel of Fig.~\ref{fig:chi_5AU}). Note the extended dead zone when the grains are small (right column). In this scenario, severe magnetic diffusivity prevents the magnetic field from coupling to the gas for $z/H \lesssim 2.5$ and ambipolar diffusion dominates at higher $z$ where the MRI grows.}
\label{fig:grains1_1}
\end{figure*}

Figs.~\ref{fig:grains1} and \ref{fig:grains1_1} compare the vertical structure and growth 
rate of the most unstable MRI modes at $R = 10$ and 5 AU, respectively, for different 
choices of the 
magnetic field strength. The left column of each figure displays solutions obtained 
assuming that grains have settled out of the gas phase. The remaining columns -- from 
left to right -- show results that incorporate the effect of a different population of single-
sized dust particles of radius $a = $ 3, 1 and $0.1\ \mu$m, respectively. Note how the growth rate, 
wavenumber and range of magnetic field strengths for 
which unstable modes exist are all drastically diminished when dust grains are present. 
This is expected, given the reduction in the coupling between the neutral and ionised 
components of the fluid when dust grains (particularly if they are small) are well mixed 
with the gas. Note also that the range of field strengths for which unstable modes are found 
matches quite well with the range for which the magnetic  field is expected to couple to the fluid, as discussed in section \ref{subsec:diffusivity} (see Figs.~\ref
{fig:contour_10AU} and \ref{fig:contour_5AU} and compare the maximum magnetic field strength for which the MRI grows and the maximum $B$ that can couple to the gas at the $z$ where the modes peak).  

For the discussion that follows, it is useful to keep in mind that both the growth rate and 
the envelope of these perturbations are shaped by the interplay of different diffusion mechanisms, 
whose relative importance vary strongly with height. In particular, the ambipolar 
diffusion  component of the diffusivity tensor drives the local growth rate (and thus the
amplitude of global perturbations) to increase with $z$ (SW03). This is because, in this diffusion regime, the maximum growth rate increases with the 
local $\chi$ (W99) and is, therefore, a strong function of the vertical location. Accordingly,
the envelopes of the perturbations driven by this term typically peak at an 
intermediate height above the midplane.  On the contrary, the maximum growth rate of Hall 
perturbations is insensitive to $\chi$ (W99). Because they are not driven from any particular vertical location, their envelope is fairly flat (SW03). We now analyse the structure and growth of the perturbations at the two radii of 
interest. The properties of these modes without dust grains were 
discussed in detail in SW05. For the sake of clarity, these results are briefly summarised as part of this discussion.

At $10$ AU, ambipolar diffusion drives the MRI when the 
field is weak ($B \lesssim 10$ mG) and ions and 
electrons are the sole charge carriers (see top panel of Fig.~\ref{fig:chi_10AU}). As a result, unstable modes peak above the 
midplane in this scenario (left column of Fig.~\ref{fig:grains1}; e.g. the mode computed with $B = 1$ mG peaks at $z/H \sim 
0.5$). For stronger fields, 
ideal-MHD holds at all $z$. This explains the flat envelope, and fast growth, of the 
perturbations obtained with $B = 10$ and 100 mG. Unstable modes are found in this 
case for $B \lesssim 250$ mG and they grow at about the ideal-MHD rate for 2 mG $
\lesssim B \lesssim 50$ mG. When $B$ is even stronger than 250 mG, the wavelength of the 
most unstable mode is $\sim H$, the disc tidal scaleheight, and the perturbations are 
strongly damped (Balbus \& Hawley 1991). 
Finally, note that no dead zone develops in this scenario, given that the magnetic field is coupled to the gas even at the midplane (see top panel of 
Fig.~\ref{fig:contour_10AU}). 

We now turn our attention to the solutions obtained under the assumption that dust 
grains are present at this radius. 
Note that when the dust particles are relatively large ($a = 3$ and $1 \ \mu$m, central 
two columns of Fig.~\ref{fig:grains1}), the perturbations exhibit the flat envelope that is typical when Hall 
diffusion controls their behaviour (for these fluid conditions Hall diffusion is expected to modify MRI modes for $z/H \lesssim 2.5$, Fig.~\ref{fig:chi_10AU}). These modes grow even at the midplane, a result consistent with the level of magnetic coupling associated with this  
disc model (e.g. see middle panel of Fig.~\ref{fig:contour_10AU}). On 
the other hand, when the dust grains are small (rightmost column of Fig.~\ref
{fig:grains1}) the low magnetic 
coupling, especially within two 
scaleheights of the midplane (Fig.~\ref{fig:contour_10AU}, bottom panel), causes the 
amplitude of all perturbations in 
this section of the disc to be severely reduced. Unstable modes were found 
here only for $B \lesssim 10$ mG, a much reduced range compared with the 
$B \approx 250$ mG for which they exist when ions and electrons are 
the only charge carriers.  

Fig.~\ref{fig:grains1_1} shows the solutions obtained for $R =  5$ AU. At this 
radius, MRI unstable modes grow -- without grains -- for $B \lesssim 795$ mG. 
Moreover, they grow at essentially the ideal-MHD rate for a significant subset of this 
range ($200$ mG $\lesssim B \lesssim 500$ mG). Hall diffusion modifies the structure 
and growth of these modes for $B \lesssim 10$ mG. When the field is within this limit, a 
small magnetically dead region develops but it disappears for stronger $B$. Note that all the solutions that incorporate dust grains peak at a height above 
the midplane where ambipolar diffusion is locally dominant (e.g. $z/H \sim 4$ for $B = 10$ mG 
and $a = 1 \ \mu$m; see also the middle panel of Fig.~\ref{fig:chi_5AU}). This signals that this diffusivity term shapes the structure of these perturbations and explains why their amplitude increase with height. The solutions computed with the small grain 
population ($a = 0.1 \ \mu$m) exhibit, as in the $R = 10$ AU case, an extended dead 
zone encompassing the region where the magnetic coupling is insufficient to sustain 
the MRI ($z/H \lesssim 2.5$; see bottom panel of Fig.~\ref{fig:contour_5AU}). Solutions are found in this case for $B \lesssim 16$ mG. 

\begin{figure*} 
\centering
 \includegraphics[width=0.8\textwidth]{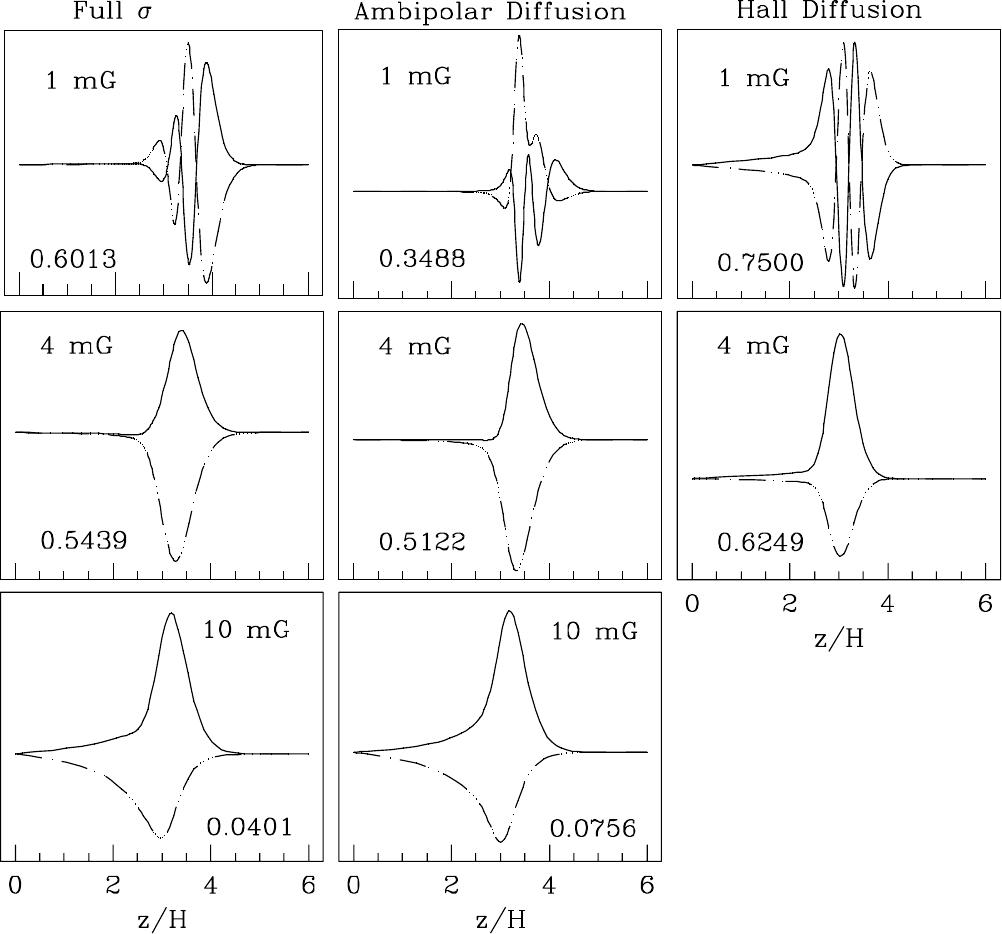} 
\caption{Structure and growth rate of the fastest growing MRI modes as a function 
of height for $R = 10$ AU and assuming $0.1 \ \mu$m grains are present. Different 
configurations of the diffusivity tensor are shown. The left column displays solutions 
incorporating all $\bmath{\eta}$ components ($\etaA$, $\etaH$ and $\etaO$). The middle and right columns correspond to the ambipolar 
($\etaH = 0$) and Hall diffusion ($\etaA = 0$) limits, respectively. We find that for relatively weak fields ($B = 1$ and 4 
mG), Hall diffusion causes the perturbations to grow faster and closer to the midplane 
than solutions in the ambipolar diffusion limit. For $B = 10$ mG, the structure of the modes computed with and without the Hall term are fairly similar, as this diffusion mechanism is no longer important at the $z$ where they peak (see bottom panel of Fig.~\ref{fig:contour_10AU}).}
\label{fig:grains3_1}
\end{figure*}

The solutions described so far in this section incorporate all diffusion 
mechanisms (represented by $\etaA$, $\etaH$ and $\etaO$). For 
comparison,  Fig \ref{fig:grains3_1}  displays how the full $\bmath{\eta}$ modes at 10 AU and including
$0.1 \ \mu$m-sized grains (left column), are modified if only ambipolar diffusion (middle column) and Hall diffusion (right column) are considered. Note 
that when the field is weak (e.g. the solutions for $B = 1$ and 4 mG), full $\bmath{\eta}$ 
perturbations grow faster, and are 
active closer to the midplane, than modes obtained in the ambipolar diffusion 
limit. This is the result of the contribution of the Hall diffusivity term (e.g. note that the solution in the Hall limit for $B = 1$ mG grows at the ideal-MHD rate). For $B = 10$ mG, the structure of the modes computed with and without the Hall term are fairly similar, as this diffusion mechanism is no longer important at the $z$ where they peak (see bottom panel of Fig.~\ref{fig:contour_10AU}). Note also that modes in the Hall limit do not grow for $B > 4$ mG, a result consistent with ambipolar diffusion being dominant at these field strengths (see bottom panel of Fig.~\ref{fig:contour_10AU}). 

\begin{figure*} 
\centering
 \includegraphics[width=0.75\textwidth]{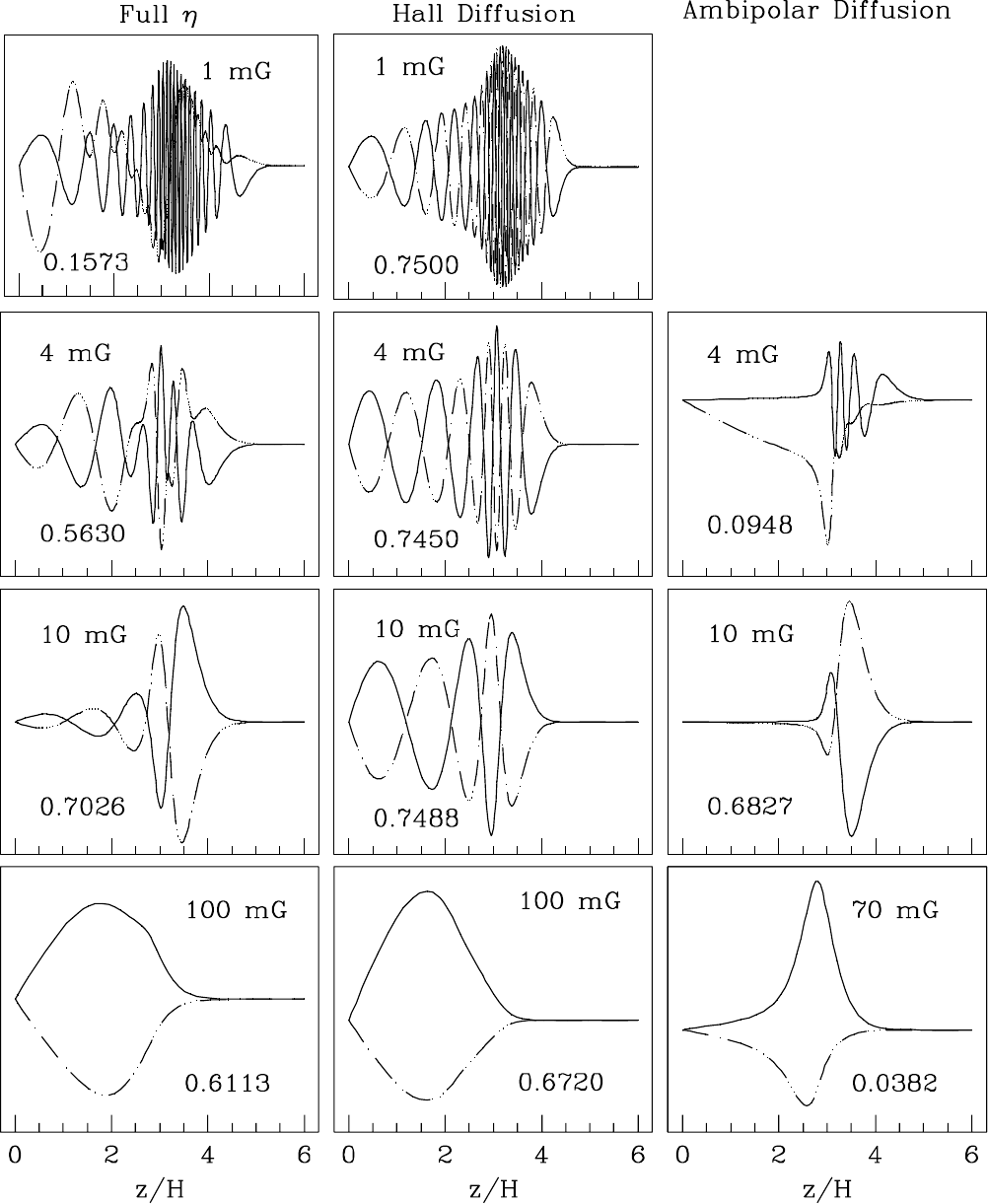} 
\caption{Structure and growth of the fastest growing MRI modes as a function 
of height at $R = 5$ AU and assuming $3 \ \mu$m grains are present. The left column 
shows solutions incorporating all $\bmath{\eta}$ components ($\etaA$, $\etaH$ and $\etaO$). The middle and right columns 
display the Hall ($\etaA = 0$) and ambipolar diffusion ($\etaH = 0$) limits, respectively.  
Hall diffusion strongly modifies the structure and growth 
of unstable modes. }
\label{fig:grains5}
\end{figure*}

A similar comparison is shown in Fig.~\ref{fig:grains5} for $R = 5$ AU and incorporating 
the effect of $3 \ \mu$m-sized dust particles. The Hall term is 
important here also, as evidenced by the faster growth and more extended unstable zone of the perturbations that include this diffusion 
mechanism. Ambipolar 
diffusion strongly influences the structure of the modes
for intermediate field strengths (e.g. compare the solutions with $B = 10$ mG and different configurations of the 
diffusivity tensor) and the perturbations 
grow in this limit for a reduced range of $B$ in relation to that associated with
full $\bmath{\eta}$ (or Hall limit) modes.

\begin{figure} 
\centering
 \includegraphics[width=0.45\textwidth]{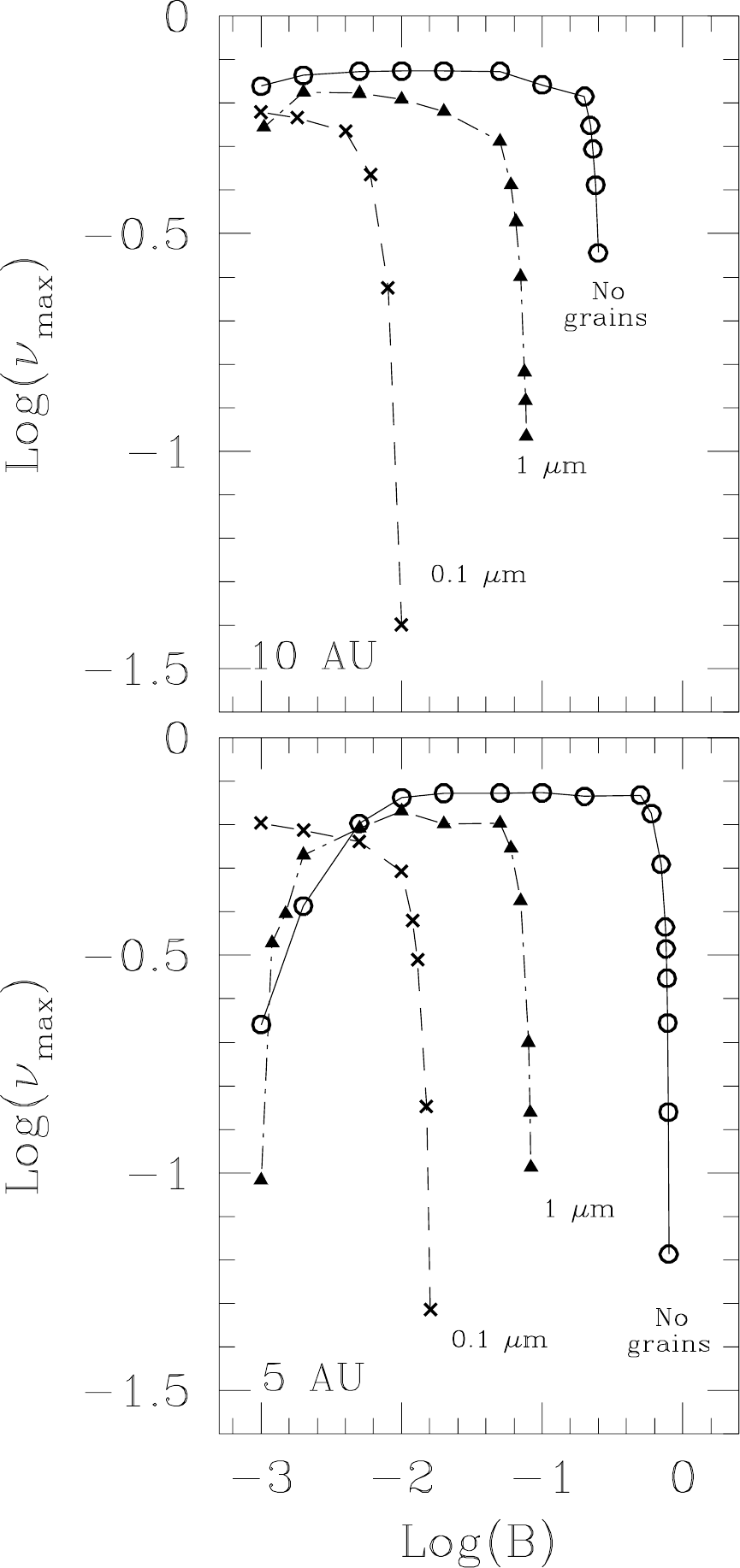} 
\caption{Growth rate of the most unstable MRI modes as a function 
of the strength of the magnetic field for different assumptions regarding the presence, 
and size, of dust grains mixed with the gas. Solutions are presented for $R = 10$ AU (top 
panel) and $R = 5$ AU (bottom panel).}
\label{fig:grains6}
\end{figure}

The dependence of the growth rate of the most unstable mode ($\nu_{\rm max}$) with 
the strength of the magnetic field is summarised in Fig.~\ref{fig:grains6} for different 
assumptions regarding the presence, and radius, of dust grains mixed with the gas. Results are shown for $R = 
10$ AU (top panel) and 5 AU (bottom panel). Note that $\nu_{\rm max}$ drops sharply at 
a characteristic field strength ($B_{\rm max}$), which is a function of both the properties 
of the grain population and the radial position. The maximum field strength for 
which perturbations grow in a particular scenario is weaker at larger radii, an effect that is particularly noticeable when no grains are 
present.  This behaviour results from the instability being (generally) damped when the midplane ratio of the Alfv\'en to sound speed approaches unity (e.g. $v_{\rm A0}/c_{\rm s} \sim 1$; Balbus \& Hawley 1991)\footnote{Note, however, that perturbations computed in the Hall limit have been found to grow for $v_{\rm A0}/ c_{\rm s} \lesssim 3$ (SW03) when the magnetic field is counteraligned with the disc angular velocity vector ($\bmath{\Omega}$).} and the wavelength of the most unstable mode becomes $\sim H$. As the midplane density and temperature decrease with radius, the ratio $v_{\rm A0}/
c_{\rm s}$, associated with a particular field strength, increases at larger radii and the perturbations are damped at a weaker field as $r$ increases. Note also that for each radii, the range of field strengths over which unstable modes 
exist is smaller as the grain size diminishes. This is consistent with the drop in 
magnetic coupling for a particular field strength as the dust particles are smaller (see 
Figs. \ref{fig:contour_10AU} and \ref{fig:contour_5AU}).

 Finally, note that at 5 AU and for weak fields ($B \lesssim 5$ mG, bottom panel of Fig.~\ref{fig:grains6}) the MRI grows faster  when the small grains are considered than it does in the other scenarios. This is because in this case the modes grow high above the midplane, where the magnetic coupling is more favourable, and are completely suppressed at lower $z$. In the other cases, on the contrary, the perturbations grow over a more extended section of the disc and the comparatively low magnetic coupling closer to the midplane reduces their global growth rate. This effect is not so noticeable at 10 AU by the relatively high magnetic coupling even at $z = 0$ at this radius.

\section{Discussion}
	\label{sec4:discussion}

In this 
paper we have examined illustrative examples of the impact of dust 
grains in the magnetic activity of protoplanetary discs and, in particular, in the linear growth and vertical structure of MRI perturbations. Solutions were 
computed for $R = 10$ and $5$ 
AU assuming a single size grain population of radius $a = 0.1$, 1 or 3 $\mu$m, and constituting 1 \% of the total mass of the gas, is well 
mixed with the gas phase over the entire vertical extent of the disc. This fraction is independent of height, so we have also assumed that the grains have not sedimented towards the midplane.
Our results indicate that the perturbations' wavenumber and growth rate are 
significantly reduced when grains are present. Furthermore, the magnetically inactive 
-- dead -- zone, 
which was practically non-existent when grains were settled, extends to 
$z/H \sim 3$ at either radii when 0.1 $\mu$m-sized grains are considered.
At 10 AU (5 AU), unstable perturbations were found in this case for $B \lesssim 10$ mG (16 mG), a much reduced range compared with the strengths for which they exist when no grains are involved (250 and 795 mG, respectively). 
This maximum field strength corresponds well to 
the equipartition field at 
the height at which the 
perturbations peak, 
as expected (e.g. $z/H \approx 3.7$ at 10 AU; see lower right panel of Fig.~\ref{fig:grains1}). 

These results illustrate the impact of 
dust particles on the dynamics -- and evolution -- of low conductivity discs. They can also be used to estimate the maximum magnetic 
field strength to support magnetic activity at $1$ AU. The magnetic coupling 
at this radius, in a disc including $0.1\  \mu$m grains, is too weak below 
$z/H \sim 3.5$ to allow the field to sufficiently couple with the gas
(W07). Assuming 
that the maximum field strength for the MRI to grow is also of the order of the 
equipartition field at about this height, we can roughly 
estimate that the MRI 
should be active for $B \lesssim 400$ mG at $1$ AU. This is a smaller range 
than the several gauss for which unstable modes exist when dust 
grains are not present (SW05). However, MRI modes still grow in this case for a wide 
range of field strengths.

The results just presented were obtained assuming that all grains have the 
same size and are well mixed with the gas at all $z$. More realistic spatial,
 and size, distributions must incorporate the effects of dust dynamics and 
evolution within the disc. Observations of the 
mid -- and far -- infrared spectra of 
discs have provided credible indications 
that dust properties in discs are indeed different from those of 
particles in diffuse clouds (e.g. Van Dishoeck 2004, D'Alessio et al. 2006 and references therein). 
Two aspects of this dust evolution have been clearly identified. First, dust 
grains coagulate from $0.1 \ \mu$m to 
$\sim 1$ mm particles. Second, 
(silicate) 
material becomes crystallised. It is believed that this crystallisation 
occurs in the disc, given that crystalline silicates are absent from the 
interstellar medium. Furthermore, the presence of this material at 
radial locations where the temperature is too low to produce them, suggests 
significant radial mixing takes place as well (e.g. Van Dishoeck 2004 and 
references therein). 
Finally, simulations of dust dynamics and evolution also suggest that 
in quiescent environments, the grains tend to settle and agglomerate 
into bigger particles (e.g. Weidenschilling \& Cuzzi 1993) and efficiently 
coagulate and grow icy mantles (Ossenkopf 1993). All these processes modify the 
surface area of dust grains and impact the recombination rate on their 
surfaces and the way they drift in response to magnetic stresses. It is also expected that a residual population of small grains would remain (e.g. Nomura et al. 2006) even when the mean grain size maybe relatively large ($a \gtrsim 1 \mu$m). This is an important consideration because these small grains tend to carry a significant fraction of the grain charge (S07). The effect of this settling 
in 
the spectral energy distribution (and optical appearance) of protostellar 
discs has been investigated in recent studies (e.g. Dullemond \& Dominik 
2004). 

How quickly, and to what 
height, dust particles are able to settle is an important, and largely 
unanswered, question. 
According to Nakagawa, Nakazawa \& Hayashi (1981), the mass fraction of 
$\sim 1$ - $10 \ \mu$m grains well mixed with the gas, diminishes from 
$\sim 10^{-1}$ to $10^{-4}$ in a timescale of about $2 \ee 3 $ to $10^{5}$ 
years. Moreover, although the timescale for dust grains to sediment  all the 
way to the 
midplane may exceed the lifetime of the disc, they may be able to settle 
within a few scaleheights from the midplane in a shorter timescale 
(Dullemond \& Dominik 2004). This 
is complicated even more by the expectation that the transition between 
sections were dust grains are well mixed with the gas, and those completely 
depleted of them, occurs gradually (Dullemond \& Dominik 2004).

MHD turbulence may itself be an important factor for the settling of dust 
particles. It may, in particular, produce sufficient vertical stirring to 
prevent settling below a certain height (Dullemond \& Dominik 2004, Carballido et al. 2005, Turner et al. 2006). However, this is contingent on the disc being able to 
generate and sustain MHD turbulence in the vertical sections where the dust is 
present. This is not guaranteed, even if turbulence exists in other regions, 
as dust grains efficiently reduce the ionisation fraction (and magnetic coupling) of the gas. As a result, the 
efficiency -- and even the viability -- of MHD turbulence in the presence of dust 
grains, is an important topic that merits careful investigation.

\section{Summary}
\label{sec:summary}

We have explored in this paper the linear growth and vertical structure of MRI unstable modes at two representative radii ($R = 5$ and 10 AU) in protoplanetary discs.  Dust grains are assumed to be well mixed with the fluid over the entire section of the disc and, for simplicity, are taken to have the same radius ($a = 3$, 1 or 0.1 $\mu$m). They constitute a constant fraction (1\%) of the total mass of the gas. These solutions are compared with those arrived at assuming that the grains have settled to the midplane of the disc (SW05). We have also explored which disc sections are expected to be magnetically coupled and the dominant diffusion mechanism as a function of height and the strength of the magnetic field, which is initially vertical. 

Our models use a minimum-mass solar nebula disc (Hayashi 1981, Hayashi et al. 1985) and incorporate all three diffusion mechanisms between the magnetic field and the neutral gas: Ohmic, Hall and Ambipolar. The diffusivity components are a function of height (as is the density) and are obtained using the method described in W07, to which we refer the reader for details. Essentially, this formalism uses a chemical reaction scheme similar to that of Nishi, Nakano \& Umebayashi (1991), but it incorporates higher dust-grain charge states that are likely to occur in discs on account of their larger gas density and temperature in relation to those of molecular clouds. Our calculations also include a realistic ionisation profile, with the main ionising sources being cosmic rays, X-rays and (to a lesser extent) radioactive decay.

Solutions were obtained at the two radii of interest for different grain sizes and configurations of the diffusivity tensor as a function of the magnetic field strength. We refer the reader to SW03 and SW05 for further details of the integration procedure. The main findings of this study are detailed below.

\subsubsection*{Magnetic diffusivity}
\begin{enumerate}
\item When no grain are present, or they are $\gtrsim 1\ \mu$m in radius, the midplane of the disc remains magnetically coupled for field strengths up to a few gauss at both radii.
\item In contrast, when a population of small grains ($a = 0.1 \mu$m) is mixed with the gas, the section of the disc below $z/H \sim 2$ ($z/H \sim 2.5$) is magnetically inactive at $R = 10$ AU (5 AU). Only magnetic fields weaker than 25 mG (50 mG) can couple to the gas.
\item At 5 AU, Ohmic diffusion dominates for $z/H \lesssim 1.2$ when the field is relatively weak ($B \lesssim$ a few milligauss), irrespective of the properties of the grain population. Conversely, at 10 AU this diffusion term is unimportant in all the scenarios studied here.
\item High above the midplane ($z/H \gtrsim 4.5 - 5$, depending on the specific model), ambipolar diffusion is severe and prevents the field from coupling to the gas for all $B$. This is consistent with previous results by W07.
\item Hall diffusion is dominant for a wide range of field strengths and grain sizes at both radii (see Figs.~\ref{fig:contour_10AU} and \ref{fig:contour_5AU}).
\end{enumerate}

\subsubsection*{Magnetorotational instability}
\begin{enumerate}
\item The growth rate, wavenumber and range of magnetic field strengths for which unstable modes exist are all drastically diminished when dust grains are present, particularly when they are small ($a \sim 0.1 \ \mu$m; see Figs.~\ref{fig:grains1} and \ref{fig:grains1_1}).
\item In all cases that involve dust grains, perturbations that incorporate Hall diffusion grow faster than those obtained under the ambipolar diffusion approximation.
\item At 10 AU, unstable MRI modes grow for $B \lesssim 80$ mG (10 mG) when the grain size is $1 \ \mu$m ($0.1 \ \mu$m), a much reduced range compared with the $\sim 250$ mG for which they exist when ions and electrons are the only charge carriers (SW05). When the grains are relatively large ($a = 1$ and $3 \ \mu$m), Hall diffusion controls the structure the modes, which grow even at the midplane. In contrast, when the grains are small ($a = 0.1 \ \mu$m), the perturbations grow only for $z/H \gtrsim 3$ and are shaped mainly by ambipolar diffusion. 
\item At 5 AU, MRI perturbations exist for $B \lesssim 80$ mG (16 mG) when the grains are $1 \ \mu$m ($0.1 \ \mu$m) in size. For comparison, the upper limit when no grains are present is $\sim 800$ mG (SW05). These modes are shaped largely by ambipolar diffusion, the dominant mechanism at the height where they peak.
\end{enumerate}

We conclude that in protoplanetary discs, the magnetic field is able to couple to the gas and shear over a wide range of fluid conditions even when small dust grains are well mixed with the gas. Despite the low magnetic coupling, MRI modes grow for an extended range of magnetic field strengths and Hall diffusion largely determines the properties of the perturbations in the inner regions of the disc.
 
\section*{ACKNOWLEDGMENTS}
\label{sec:Acknowledgments}

This research has been supported by the Australian Research Council.
RS acknowledges partial 
support from NASA Theoretical Astrophysics Program Grant NNG04G178G.

\bsp
\label{lastpage}
\end{document}